\documentclass{emulateapj}

\begin{document}
\title{PORTRAIT OF A DARK HORSE: A PHOTOMETRIC AND SPECTROSCOPIC STUDY OF THE ULTRA-FAINT MILKY WAY SATELLITE PEGASUS\,III}
\author{Dongwon Kim,$\!$\altaffilmark{1} Helmut Jerjen, $\!$\altaffilmark{1} Marla Geha,$\!$\altaffilmark{2}  Anirudh Chiti,$\!$\altaffilmark{3} Antonino P. Milone,$\!$\altaffilmark{1} Gary Da Costa,$\!$\altaffilmark{1} Dougal Mackey,$\!$\altaffilmark{1} Anna Frebel $\!$\altaffilmark{3}, and Blair Conn $\!$\altaffilmark{1}}

\altaffiltext{1}{Research School of Astronomy and Astrophysics, Australian National University, Canberra, ACT 2611, Australia}
\altaffiltext{2}{Astronomy Department, Yale University, P.O. Box 208101, New Haven, CT 06510, USA}
\altaffiltext{3}{Department of Physics and Kavli Institute for Astrophysics and Space Research, Massachusetts Institute of Technology, Cambridge, MA 02139, USA}

\email{dongwon.kim@anu.edu.au}

\begin{abstract}
Pegasus\,III (Peg III) is one of the few known ultra-faint stellar systems in the outer halo of the Milky Way. 
We present the results from a follow-up campaign with Magellan/IMACS and Keck/DEIMOS. 
Deep stellar photometry down to $r_0\approx 25$\,mag at 50\% completeness level has allowed accurate measurements of its photometric and structural properties. The color-magnitude diagram of Peg III confirms that the stellar system is well described by an old ($\gtrsim12$ Gyr) and metal-poor ([Fe/H]$\lesssim-2.0$ dex) stellar population at a heliocentric distance of $215\pm12$\,kpc.
The revised half-light radius $r_{h}=53\pm14$\,pc, ellipticity $\epsilon=0.38^{+0.22}_{-0.38}$, and total luminosity $M_{V}=-3.4\pm0.4$ are in good agreement with the values quoted in our previous paper. We further report on the spectroscopic identification of seven, possibly eight member stars of Peg\,III. The Ca II triplet lines of the brightest member stars indicate that Peg\,III contains stars with metallicity as low as [Fe/H]=$-2.55\pm0.15$ dex. Peg\,III has a systemic velocity of  $-222.9 \pm 2.6$\,km s$^{-1}$ and a velocity dispersion of $5.4^{+3.0}_{-2.5}$\,km s$^{-1}$.
The inferred dynamical mass within the half-light radius is $1.4^{+3.0}_{-1.1} \times 10^6\rm{M_{\odot}}$ and the mass-to-light ratio $\textup{M/L}_{V} = 1470^{+5660}_{-1240}$ $\rm{M_{\odot}/L_{\odot}}$, providing further evidence that Peg\,III is a dwarf galaxy satellite. We find that Peg\,III and another distant dwarf satellite Pisces\,II lie relatively close to each other ($\Delta d_{spatial}=43\pm19$\,kpc) and share similar radial velocities in the Galactic standard-of-rest frame ($\Delta v_{GSR}=12.3\pm3.7$ km s$^{-1}$). This suggests that they may share a common origin.

\end{abstract}


\altaffiltext{*}{This paper includes data gathered with the 6.5 meter Magellan Telescopes located at Las Campanas Observatory, Chile.}

\section{Introduction}
The census of satellite galaxies and star clusters associated with the Milky Way (MW) has been continuously updated for the last decade. 
Following the success of the Sloan Digital Sky Survey~\citep[SDSS;][]{York2000} ,which revealed the presence of ``ultra-faint'' ($M_V>-5$) MW satellites~\citep[e.g.][]{Willman2005,Zucker2006,Belokurov2006,Irwin2007,Walsh2007,Koposov2007,Balbinot2013,KimJerjen2015a}, recent wide-field photometric 
surveys have been instrumental in finding many more such systems in the MW halo, and probing to increasingly faint levels~\citep[e.g.][]{Bechtol2015, DWagner2015, Kim2015a, KimJerjen2015b,Kim2016,Koposov2015a,Laevens2015a,Laevens2015b,Luque2016, Martin2015,Torrealba2016a,Torrealba2016b}. A growing number of the newly discovered MW satellites are filling the gap between the classical dwarf galaxies and globular clusters in the size-luminosity plane, meaning that it is increasingly difficult to classify these systems using only these two parameters~\citep{Willman2012}. Instead, the
approach of determining their kinematics or chemistry still remains valid as a main diagnostic for distinguishing the two types of stellar 
systems~\citep[e.g. see discussions in ][]{Laevens2014,Belokurov2014,Kirby2015a,Weisz2016,Voggel2016}. Spectroscopic follow-ups 
for the kinematic and chemical properties are rapidly catching up with the discoveries of the new satellites, but it is a technical challenge to study more than a handful of member stars in these systems due to their intrinsic low total luminosities and therefore lack of bright red giant branch stars~\citep{Simon2015,Walker2015,Walker2016,Koposov2015b,Kirby2015b,Martin2015,martin2016a,martin2016b,Ji2016,Raoederer2016}.

Pegasus III (Peg III hereafter) is a MW satellite galaxy originally found in the SDSS Data Release 10 photometry~\citep{Ahn2014} by~\cite{Kim2015b}, who also provided detection confirmation at the $\sim10\sigma$ level based on DECam photometry. The follow-up imaging with DECam further revealed the presence of six blue-horizontal-branch (BHB) candidate stars. Their apparent magnitudes implied that Peg III is located at a heliocentric distance of $205\pm20$\,kpc in the outer region of the MW halo. From the relatively shallow DECam photometry, Peg III appeared to be elongated with a rather irregular stellar distribution possibly indicative of tidal disturbance. Deeper imaging was thus required to confirm its true nature.

Peg III is a member of the small population of presently known MW satellites in the distance range $130$\,kpc$< d_{GC}<250$\,kpc. It is also located close to another distant satellite, Pisces II~\citep[Psc II hereafter, $d_\odot\sim180$\,kpc;][]{Belokurov2010}. These two satellites seem to form a physical pair with an angular separation of $8.5^\circ$ on the sky and a relatively small difference in line-of-sight distance of $\sim30$\,kpc. Other close pairs of MW satellites have been reported before -- for example, Bo\"otes I \citep{Belokurov2006} and Bo\"otes  II \citep{Walsh2007}, Leo IV \citep{Belokurov2007} and Leo V \citep{Belokurov2008}, or Horologium I 
\citep{Koposov2015a,Bechtol2015} and Horologium II \citep{KimJerjen2015b}, leading to speculations about their companionship or common origin. The most notable 
pair is Leo IV - V, another pair of distant satellites ($d_\odot>150$\,kpc), for which the systemic line-of-sight velocities differ only by $\sim40$\,km\,s$^{-1}$~\citep{Simon2007,Belokurov2008}, supporting the scenario that the 
pair might be gravitationally bound as a ``tumbling pair" \citep{deJong2010}. In this context, the discovery of another close pair of distant MW satellites naturally raises the question as to whether 
their systemic velocities are also similar to each other. To find an answer requires spectroscopic follow-up to obtain their kinematic information.

We observed Peg III with Magellan/IMACS for deep photometry and Keck/DEIMOS for spectroscopy in order to firmly establish its 
luminosity and structural parameters, estimate its dynamic mass-to-light ratio and investigate its possible association with 
Psc II. 

\section{Photometry and Astrometry}

\begin{figure*}
\begin{centering}
\includegraphics[scale=1]{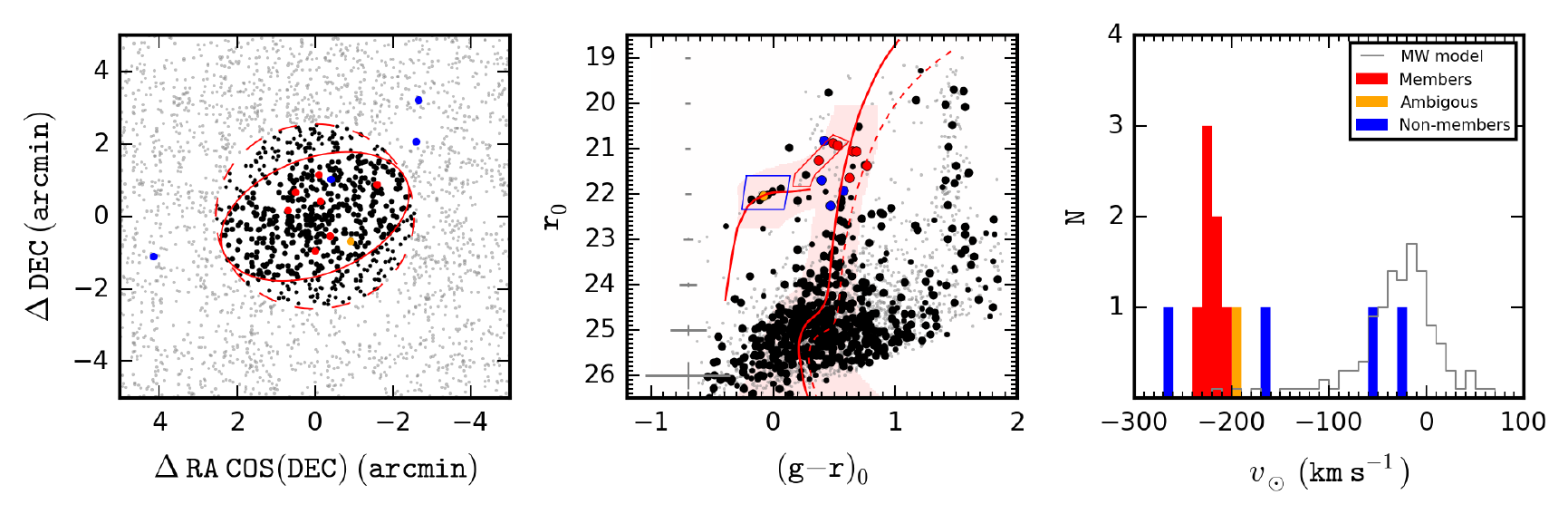}
\par\end{centering}
\caption{Left panel: distribution of all objects classified as stars in a $10\farcm0 \times10\farcm0$ field centred on Peg III. Large black dots are the stars within an ellptical radius of $2\farcm55$, equivalent to $3$ half-light radii, of the center of Peg\,III (red ellpse) whereas small dots the stars outside the ellipse but within a circular-radial distance of $2\farcm55$ (dashed circle). The red, blue and orange large dots represent the 12 stars for which we obtained velocity measurements with Keck/DEIMOS, where red (blue) dots identify kinematic members (non-members). The small gray dots are all the rest of stars from our IMACS photometry. The orange large dot is a star whose membership is ambiguous. Middle panel: Magellan/IMACS CMD of the stars in the left panel. The symbols are the same as in the left panel. The two Dartmouth isochrones~\citep{Dartmouth} of age 13.5 Gyr, [Fe/H]=$-2.5$ and [$\alpha$/Fe]=+0.4 (solid) and of age 12 Gyr, [Fe/H]=$-1.1$ [$\alpha$/Fe]=$+0.2$ (dashed) are overplotted at a distance of 215 kpc. The HB fiducial track has been derived from \cite{Bernard2014} by using the observed CMD of the globular cluster M\,15 ([Fe/H]=$-2.42$). The blue and red polygons highlight the HB and AGB/RHB candidate stars of Peg\,III, respectively. Right panel: radial velocity distribution of the 12 stars observed with Keck/DEIMOS. The colors are the same as in the left/middle panels. The solid line illustrates the predicted distribution of MW stars from the Besancon model~\citep{Robin2003}, within a radius of $5\farcm0$, normalized to the number of observed stars.}
\label{fig:CMD}
\end{figure*}

\begin{figure*}
\begin{centering}
\includegraphics[scale=0.75]{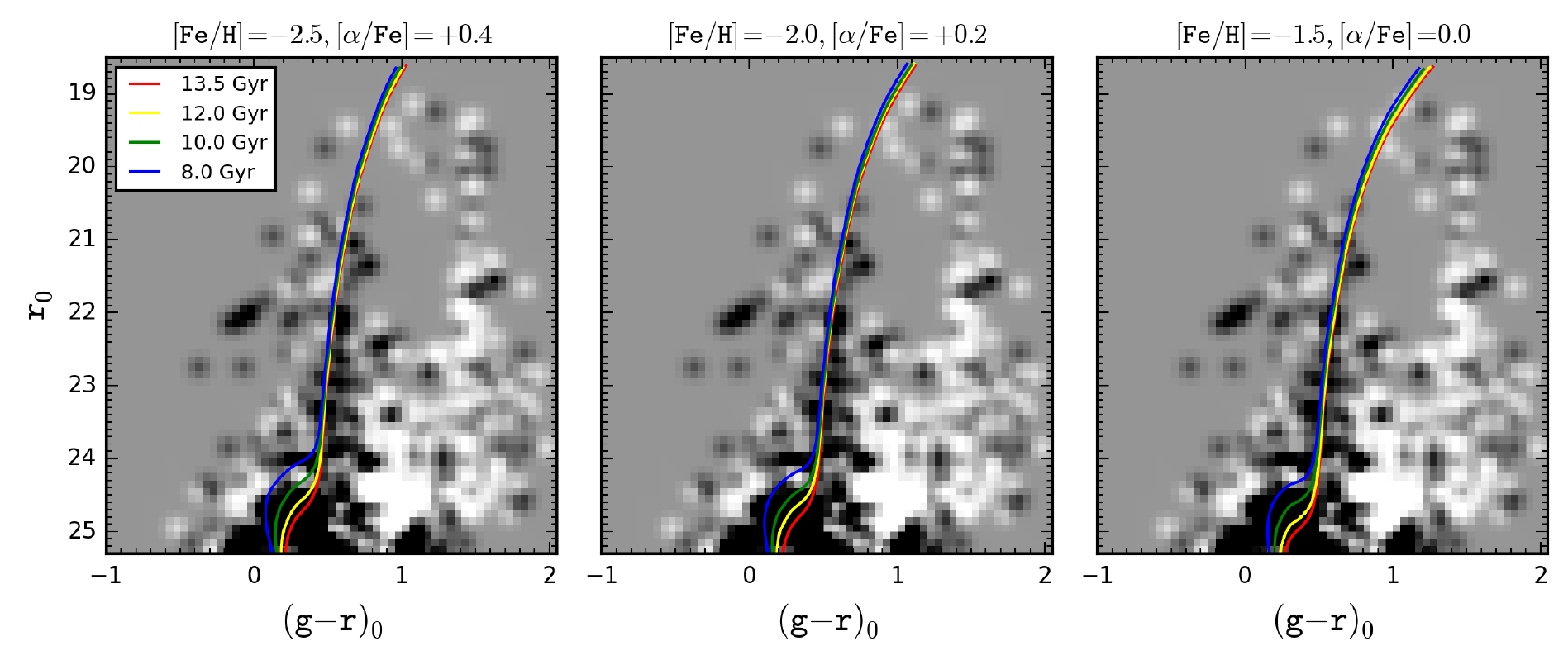}
\par\end{centering}
\caption{Background-subtracted Hess diagrams of Peg III within $2\farcm55$ (dashed circle in the left panel of Figure~\ref{fig:CMD}). Overplotted are Dartmouth isochrones at a distance modulus of 21.66 mag for different ages, metallicities, and $\alpha$ abundances.}
\label{fig:HessDiagram}
\end{figure*}

\begin{figure}[t!]
\includegraphics[scale=0.65]{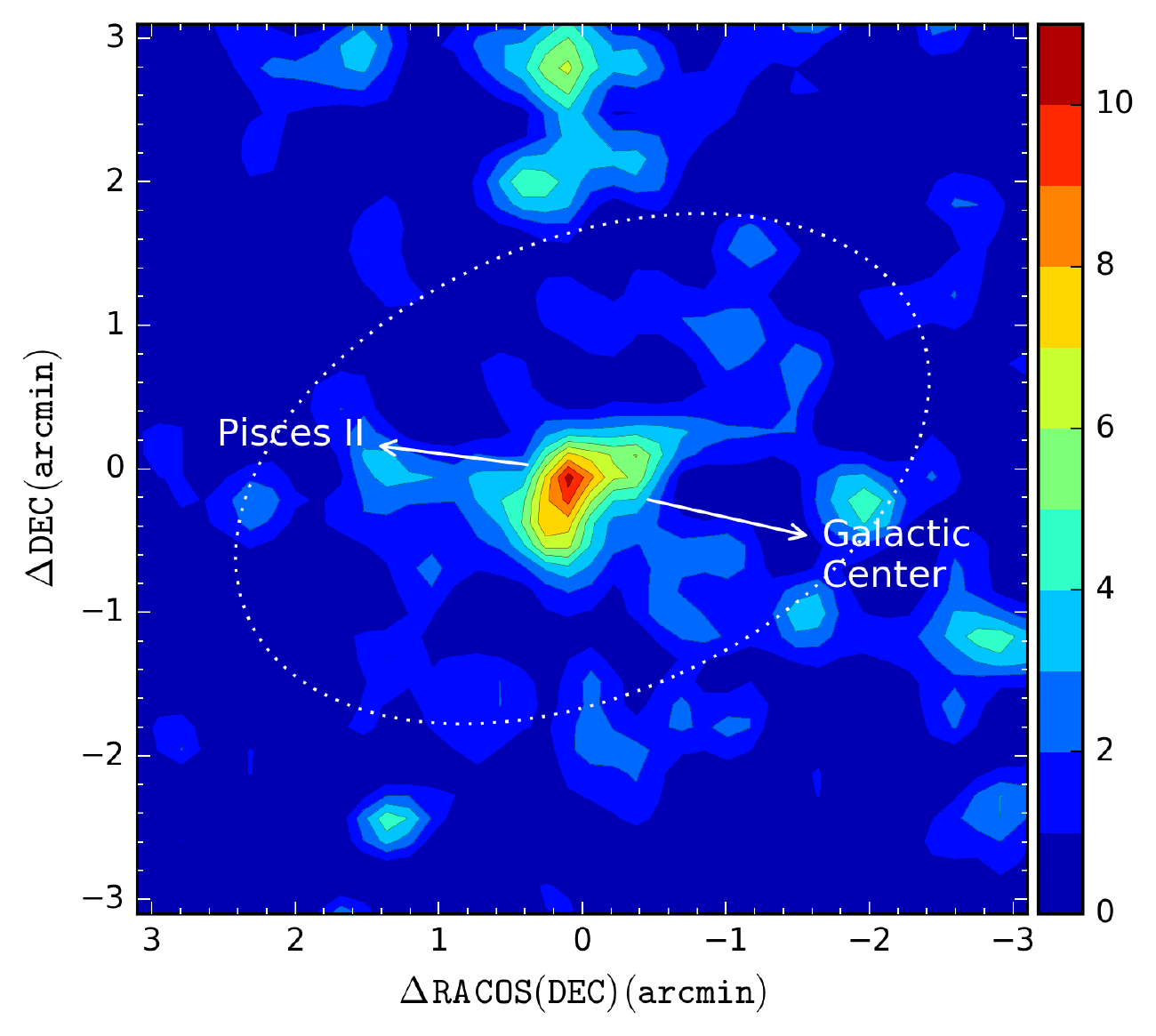}
\caption{Convolved density contour map of Peg\,III candidate stars that pass the photometric filter illustrated in the middle panel of Figure~\ref{fig:CMD}. The density map was binned with a pixel size of $10\farcs0$ and smoothed with a Gaussian kernel with FWHM of $23\farcs6$. The contours mark the levels of star density in units of the standard deviation above the background (median value). The white dotted ellipse represents $3$ half-light radii of the center of Peg\,III. The left and right arrows point to the nearby outer halo satellite Psc II and the Galactic Center, respectively. 
\label{fig:Contour}}
\end{figure}

We observed Peg III on 2015 July 22nd with the f$/4$ mode of the Inamori-Magellan Areal Camera \& Spectrograph (IMACS) at the Magellan/Baade Telescope in the $g$ and $r$ bands. Magellan/IMACS has eight 2k$\times$4k CCDs with a total field of view of $15\farcm4\times15\farcm4$ and a pixel scale of $0\farcs2$ pixel$^{-1}$ ($2\times2$ binning). 

We obtained a series of $17\times600$s dithered exposures in $g$ and $15\times600$s in $r$ together with 20 bias frames, 10 dome flats in each filter taken before the science exposures, and 7 sky flats for each filter taken at the end of our observing night. During the observing run, the weather was clear and seeing ranged from $0\farcs8$ to $1\farcs2$. We produced the master bias and master flats using the zerocombine and flatcombine tasks in IRAF, and then carried out bias subtraction and flat fielding using the imarith task. 

To find the astrometric solutions for the reduced science images, we used SCAMP~\citep{Scamp} and the SDSS DR 10 photometry catalog \footnote{http://www.sdss3.org/dr10/}. We then combined the reduced images into our final image stacks using SWARP~\citep{Swarp}.

We performed point-spread function (PSF) photometry on the final image stacks using SExtractor/PSFex~\citep{SExtractor,PSFEx}. 
These software programs provide the $\mathtt{SPREAD\_MODEL}$ parameter that allows morphological star/galaxy separation, for which we set a threshold $\mathtt{\left| SPREAD\_MODEL\right|<0.003+SPREADERR\_MODEL}$~\citep[see e.g.][]{Desai2012,Koposov2015a}. This selection process was applied to the $g$ band image stack, which has a longer total integration time than the $r$ band image stack. After crossmatching the $g$ and $r$ catalogs using STILTS~\citep{STILTS} with a $1\arcsec$ tolerance, we converted the instrumental magnitudes of the matched catalog into the SDSS photometric system using unsaturated stars in common with our 
DECam photometry catalog for Peg III presented in~\cite{Kim2015b}, via bootstrap sampling with 500 iterations and 3-sigma clipping. Finally, we corrected the calibrated magnitudes for Galactic extinction based on the reddening map by~\cite{Schlegel1998} and the correction coefficients from~\cite{Schlafly2011}. In the studied field of view, the typical value of E(B-V) is $\sim0.124$.

We note that the magnitudes of seven objects in the star catalog were replaced by average measurements from two best-seeing individual exposures as they fell onto the edges or corners of CCD chips in some individual exposures and suffered the extra-widening of the PSF relative to the typical full width half maximum (FWHM) in the process of image stacking. Such stacking-induced degrading of the PSF at CCD chip boundaries becomes more obvious when the individual exposures have a seeing difference as large as the pixel scale. We searched in our sample for stars brighter than $r_0=23$ mag that have been affected by the phenomenon, and found seven objects including the stars \#1 and \#8 in our spectroscopy sample (see Table~\ref{tab:TargetList}). This effectively accounts to $\sim2$\% of all objects in that magnitude range\footnote{We found the stars affected by the degrading of PSF in $g$ or $r$ band by crossmatching catalogs from the stacked image and best-seeing individual frames, and filtering the matched catalog with the following criteria: \begin{itemize} \item[-] $\mathtt{fwhm_{indi}<\overline{fwhm_{indi}}+3\sigma_{fwhm,indi}}$ \item[-] $\mathtt{fwhm_{stack}>\overline{fwhm_{stack}}+3\sigma_{fwhm,stack}}$ \item[-] $\mathtt{\left| SPREAD\_MODEL_{indi}\right|<0.003+SPREADERR\_MODEL_{indi}}$ \end{itemize} }. The number of such objects in the fainter magnitude range of $r_0>23$ mag where a typical FWHM is not well defined are not precisely determined. This phenomenon is, however, unlikely to significantly affect the rest of our analysis as the portion of the affected stars is small, the resulting magnitude difference smaller than 0.1 mag, and the width of the photometric filtering mask used in Section 3 sufficiently wide to take the effect into account.

We also measured the completeness levels of our photometry as a function of color and magnitudes using artificial stars as described in~\cite{Kim2016}. At the color $(g-r)_{0}=0.40$, the 90\% and 50\%  completeness levels correspond to $r_{0}=22.65$ and 
$r_{0}=24.92$, respectively.  

\section{Satellite Distance and Stellar Population}

The distribution of stars in our IMACS photometry and corresponding color-magnitude diagram are presented in Figure~\ref{fig:CMD}, reaching $\sim3$ mag fainter than our previous DECam photometry at the same S/N levels. The stars within an elliptical radius of $2\farcm55$, equivalent to 3 half-light radii, of the center of Peg\,III are highlighted with black large dots. The stars outside the 3 half-light radii but within a circular radius of $2\farcm55$ are also highlighted with smaller black dots to take into account the large uncertainty of ellipticity derived in Section 3. The large red and blue dots in Figure~\ref{fig:CMD} represent kinematically confirmed member and non-member stars respectively (see Section 4 for more details).

We constrain the heliocentric distance of Peg III using the luminosity of its HB and the fiducial HB track of a globular cluster. Since the absolute total luminosity of Peg III was estimated to be $-4.1\pm0.5$ in our previous work, the mean metallicity of the system is expected to be as low as [Fe/H]$\sim-2.5$ according to the mass-metallicity relation by~\cite{Kirby2013}. We note that the recent metallicity measurements of the MW satellite dwarf galaxies in the same luminosity range as Peg III, for example Psc II ~\citep[$\langle\lbrack$Fe/H$\rbrack\rangle$=$-2.45\pm0.07$;][]{Kirby2015a} and Reticulum II~\citep[$\langle\lbrack$Fe/H$\rbrack\rangle$=$-2.65\pm0.07$;][]{Simon2015} are consistent with the mass-metallicity relation. Accordingly, we adopted the fiducial HB sequence of M\,15, one of the most metal-poor globular clusters ([Fe/H]=$-2.42$), from~\cite{Bernard2014}. We converted the fiducial into the SDSS photometric system with the help of transformation equations and coefficients provided by~\cite{Tonry2012}. We took literature values of E(B$-$V)=0.11 and $(m-M)_0=15.25$~\citep{Kraft2003} to obtain the reddening-corrected fiducial HB sequence. We then expressed this fiducial HB sequence as a function of color $(g-r)_0$ by means of a 5th order polynomial regression, fit this function to the six BHB candidate stars in the blue polygon in the middle panel of Figure~\ref{fig:CMD} with the least-squares method, and derived a distance modulus of $(m-M)_0=21.66\pm0.12$. For the uncertainty in the final estimate of the distance modulus, we combined in quadrature the uncertainties associated with our calibration to our DECam photometry ($<0.01$ mag), the adopted distance modulus of M\,15~\citep[$\sim0.1$ mag; ][]{Kraft2003}, our fiducial HB sequence fit ($0.03$ mag, determined by jackknife resampling), and galactic reddening in the $r$ band ($<0.01$ mag). In addition, we took into account the systematic uncertainty associated with the metallicity-luminosity relation, for which our estimate is $\sim0.05$ mag.

In the middle panel, a Dartmouth isochrone~\citep{Dartmouth} for age 13.5 Gyr, [Fe/H]=$-2.5$, and [$\alpha$/Fe]=+0.4 (solid curve), an isochrone from the same set but for age 12 Gyr, [Fe/H]=$-1.1$, [$\alpha$/Fe]=+0.2 (dashed curve), and the M\,15 fiducial HB sequence are plotted on the CMD at a distance modulus of $(m-M)_0=21.66$, or a heliocentric distance of 215 kpc. The three kinematic member stars in the red polygon appear systematically brighter than the blue horizontal branch (BHB) and bluer than the red giant branch (RGB). An excess of such stars relative to the RGB has been noticed in the CMD of the Hercules dwarf galaxy~\citep[e.g., Figure 2 in ][]{Sand2009} and the majority of them has been identified as its asymptotic giant branch (AGB) or red horizontal branch (RHB) population by photometric and spectroscopic studies~\citep[e.g.,][]{Aden2009,Fabrizio2014}. Most likely the three Peg III member stars in the red polygon are AGB/RHB stars too. Three of the other four kinematic members of Peg III are consistent with the red giant branch (RGB) while the last one, star\#2, is almost $0.1$ mag redder. That color difference cannot be explained by photometric uncertainties alone. There are different factors that can cause a color spread in the RGB, including dispersions in metallicity and carbon abundances. The metallicity of stars in MW satellite dwarf galaxies with similar total luminosities to Peg\,III ranges largely from [Fe/H]=$-3.5$ up to [Fe/H]=$-1.0$ dex~\cite[e.g., Ursa Major II; ][]{Vargas2013}. The red star \#2 of Peg III can be indeed fitted with an isochrone for a higher metallicity of [Fe/H]=$-1.1$ (dashed curve in the middle panel) at the same distance modulus. Carbon stars ([Ca/Fe]$\gg+1.0$) in dwarf galaxies also tend to be redder than carbon-normal RGB stars due to the Bond-Neff effect~\citep{Bond1969}, as shown for instance in Figure 7 of~\cite{Kirby2015c}. The possibility of a metallicity and carbon spread among the Peg III stars can be tested once the information on the chemical abundances of the individual stars becomes available. The low signal-to-noise of our spectra does not permit a detailed analysis for the chemical abundances of the individual targets.

We present a background-subtracted Hess diagram in Figure~\ref{fig:HessDiagram}, which allows us to qualitatively assess the stellar population of Peg\,III by means of model isochrone fitting for different properties. The Hess diagram was constructed based on the CMD of all stars within the radial distance of $2\farcm55$ and subtracting a control CMD of all stars in an equal area outside $4\farcm0$. We overplot Dartmouth isochrones with different ages (8,10,12,13.5 Gyr) and metallicities [F/H] values (-2.5, -2.0, -1.5 dex). The [$\alpha$/Fe] values for the isochrones are determined based on the [Fe/H]-[$\alpha$/Fe] relation from~\cite{Vargas2013}. The distance modulus is fixed at 21.66 mag. The most notable difference among the isochrone fits is found in the main-sequence turnoff region, where the isochrones for metal poor ([Fe/H]$\lesssim-2.0$) and old ($\gtrsim12$ Gyr) stellar populations appear to be most consistent. This suggests that Peg\,III shares similar properties, i.e. low metallicities and old ages, of stellar populations with previously known ultra-faint MW satellite dwarf galaxies~\citep[e.g.,][and references therein]{Brown2014}.

\section{Structural Properties and Absolute Luminosity}

\begin{figure}[t!]
\begin{centering}
\includegraphics[scale=0.55]{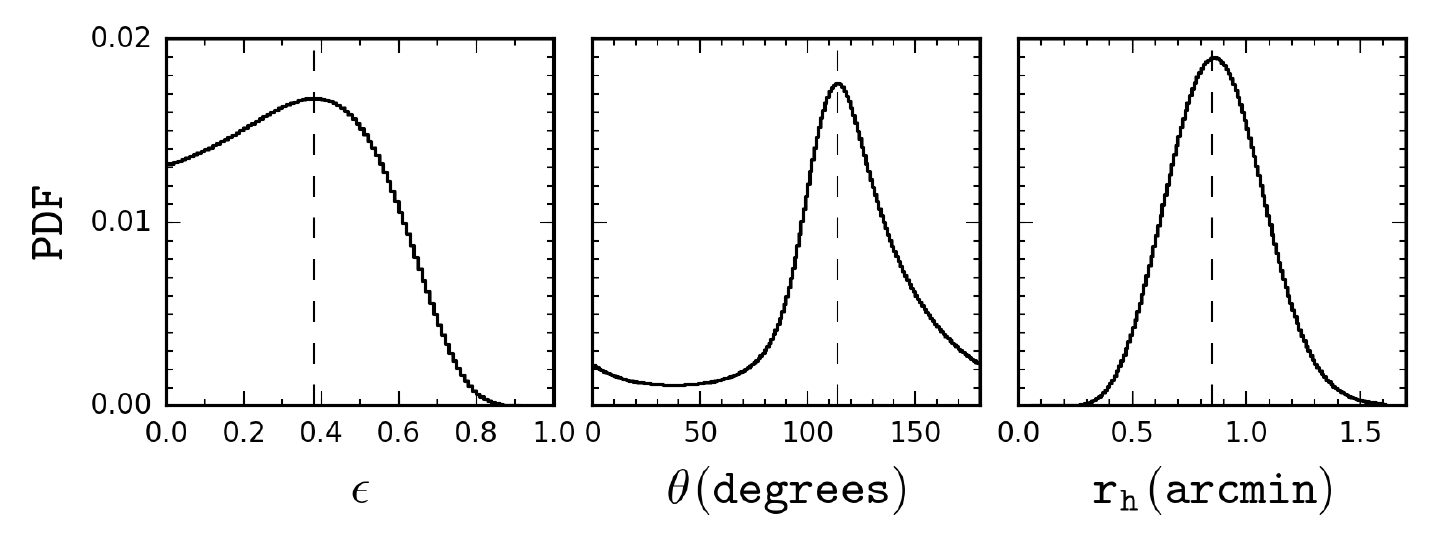}
\includegraphics[scale=0.8]{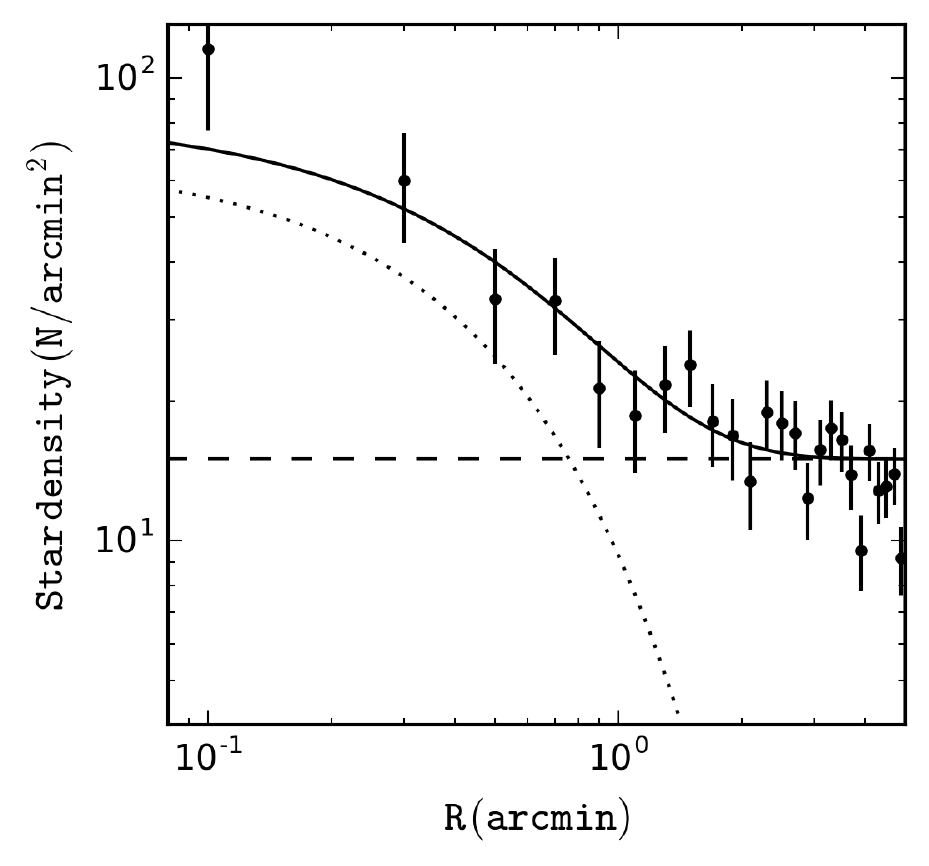}
\par\end{centering}
\protect\caption{Upper panels: marginalized probability distrubtion functions (PDFs) of the structural parameters of Peg III. Lower panel: radial stellar density profile of Peg\,III. $\textup{R}$ is the elliptical radius. Overplotted are the best exponential model based on the parameters in Table~\ref{tab:Parameters} (dotted line), the contribution of foreground stars (dashed line) and the combined fit (solid line). The error bars were derived from Poisson statistics.\label{fig:RadialProfile}}
\end{figure}

\begin{deluxetable}{lrl}
\tablewidth{0pt}
\tablecaption{Properties of Peg\,III}
\tablehead{
\colhead{Parameter} & 
\colhead{Value} &
\colhead{Unit}}
\startdata
$\alpha_{J2000}$ & 22 24 24.48 & h m s \\
$\delta_{J2000}$ & +05 24 18.0 & $^\circ$ $\arcmin$ $\arcsec$ \\
$l$ & 69.8452 & deg\\
$b$ & $-$41.8293 & deg\\
$(m-M)_0$ & $21.66\pm0.12$ & mag \\
$d_\odot$ & $215\pm12$ & kpc \\
$r_{h}$ & $0.85\pm0.22$ & $\arcmin$ \\
& $53\pm14$ & pc \\
$\epsilon$ & $0.38^{+0.22}_{-0.38}$ & \\
$\theta$ & $114^{+19}_{-17}$ & deg \\
$M_{V}$ & $-3.4\pm0.4$ & mag \\
$\textup{L}_{V}$ & $1960\pm720$ & $\rm{L_{\odot}}$ \\
$\langle v_{\odot} \rangle$ & $-222.9 \pm 2.6$ & km s$^{-1}$ \\
$v_{GSR}$ & $-67.6$ & km s$^{-1}$ \\
$\sigma_v$ & $5.4^{+3.0}_{-2.5}$ & km s$^{-1}$ \\
$\textup{M}_{1/2}$ & $1.4^{+3.0}_{-1.1}$  & $10^6\rm{M_{\odot}}$ \\
$\textup{M/L}_{V}$ & $1470^{+5660}_{-1240}$ & $\rm{M_{\odot}/L_{\odot}}$ 
\enddata
\label{tab:Parameters}
\end{deluxetable}

Figure~\ref{fig:Contour} shows the convolved contour map of star density centred on Peg III made of stars that passed a photometric filtering mask constructed from the Dartmouth isochrone for age 13.5 Gyr, [Fe/H]=$-2.5$, and [$\alpha$/Fe]=+0.4 and the M\,15 HB fiducial line, as illustrated with a light-red shadow in Figure~\ref{fig:CMD}. The width of the mask gradually increases in the faint regime to take into account photometric uncertainties. The shape of the outer isodensity lines still remains  irregular in the deep imaging data as previously seen in our DECam data~\citep{Kim2015b}, which lends support to the scenario that the observed irregularity reflects the true structure, rather than being a consequence of the limited depth of the previous photometry. Given such a small population of stars in the system, however, assessing the observed irregularity is always subject to small number statistics~\citep[see e.g.][]{Martin2008,Walsh2008,Sand2010,Munoz2012}.

The central coordinates and structural parameters of Peg III were derived using the Maximum Likelihood (ML) routine as described in~\cite{Kim2016} based on~\cite{Martin2008} using our IMACS photometry catalog and the photometric filtering mask. The upper panels of Figure~\ref{fig:RadialProfile} show marginalized PDFs for key structural parameters. In this analysis, Peg III remains elliptical with $\epsilon=0.38^{+0.22}_{-0.38}$ at a position angle of $\theta=114^{+19}_{-17}$, but its half-light radius ($r_h=53\pm14$\,pc) appears $\sim32$\% smaller than the previous estimate~\citep[$r_h=78^{+30}_{-24}$\,pc;][]{Kim2015b}. Nevertheless the two values are consistent at the 1-sigma level. The radial density profile with the best-fit exponential model based on the resulting values is presented in the lower panel of Figure~\ref{fig:RadialProfile}. We further estimated the absolute luminosity of Peg III using the Dartmouth luminosity function (LF) for age 13.5 Gyr, [Fe/H]=$-2.5$ and [$\alpha$/Fe]=$+0.4$ with the mass function by \cite{Chabrier2001} as follows. We first normalised the theoretical LF and multiplied with the photometric completeness function derived in Section 2. We note that the Dartmouth isochrone accounts for RGB and MS stars but not AGB/HB sequences. We then integrated the scaled LF in the magnitude range of $r_0>22.0$ mag to calculate the probability density for the number of detected RGB/MS stars fainter than $r_0=22$ mag in our IMACS photometry. Accordingly, we repeated the ML routine using the previous filtering mask but excluding the AGB/HB sequences and the RGB sequence brighter than $r_0=22.0$ mag to estimate the number of RGB/MS stars fainter than $r_0=22.0$ mag that belong to the overdensity $N$ with eq. (5) in~\cite{Martin2008}. The ratio of the star count $N$ to the probability density allowed us to scale the normalised LF to the observed level. Integrating the up-scaled LF estimates the integrated total luminosity of RGB/MS stars in Peg III as $M_{r}=-3.2^{+0.3}_{-0.4}$ or $M_{V}=-3.0^{+0.3}_{-0.4}$ by luminosity weighted mean color $V-r=0.17$ mag for the model LF. Finally, we calculated the flux of AGB/HB candidate stars in the red and blue polygons presented in the middle panel of Figure~\ref{fig:CMD} using the transformation equation by~\cite{Jordi2006} to convert their $g$ and $r$ magnitudes into $V$ magnitudes. Their contribution increased the total luminosity in the $V$ band by 0.4 mag and the uncertainty by 0.1 mag for upper (fainter) limit and by less than 0.1 mag for lower (brighter) limit. Therefore, we adopted $M_{V}=-3.4\pm0.4$ as our final estimate for the total luminosity of Peg III.

 All the new estimates for the parameters are consistent with their previous estimates at the $1-\sigma$ level. The new values suggest that Peg III is somewhat smaller and fainter than previously estimated~\citep{Kim2015b}. All resulting values presented in this and the next sections are summarized in Table~\ref{tab:Parameters}.

\section{Spectroscopy}

The data were taken with the Keck~II 10-m telescope and the DEIMOS spectrograph \citep{faber2003}.  One multislit mask was observed in Peg III on the night of July 17th 2015. We selected 30 targets based on their color-magnitude distribution and distances from the center of the system using the DECam photometry from~\cite{Kim2015b}. We assigned priorities to potential RGB, AGB and HB stars selected to follow the best-fitting isochrone to the CMD of Peg\,III in the DECam data. We used the 1200~line~mm$^{-1}$\,grating that covers a wavelength range $6400-9100\mbox{\AA}$ at the spectral resolution $1.37\mbox{\AA}$ (FWHM, equivalent to 47 km s$^{-1}$ at the Ca II triplet). Slitlets were $0\farcs7$ wide. The total exposure time was 2.5 hours. 

We reduced the data using a modified version of the DEIMOS spec2d software pipeline\citep{Cooper2012,Newman2013}. We refer the reader to \citet{Simon2007} for a more detailed description of the radial velocity measurement method. We measured the spectra of 12 out of the 30 targets, and their median S/N per pixel ranged from 1.4 to 7.0.

The membership of the sample stars was determined based on radial velocity, position in the CMD and distance from the center of the dwarf galaxy.  We identify 7 secure members shown in red in Figure 1.  An eighth member (shown in Figure 1 in orange) is 30 km s$^-1$ away from the systemic velocity, but also has very large velocity errors. This star has the colors of a horizontal branch member star, and may be a RR Lyrae star. We do not include this star in the calculations below.

The velocities for all the observed Peg III candidate members are presented in Table 2. We note that the color and magnitude of star \#11 was taken from our previous DECam photometry as its PSF on the IMACS images in $g$ band was considerably affected by a saturated object nearby.

\begin{deluxetable*}{cccccccccc}

\tablewidth{0pt}
\tablecaption{Keck/DEIMOS Target List}
\tablehead{
\colhead{Object} & 
\colhead{RA (J2000)} & 
\colhead{Dec (J2000)} &
\colhead{$(g-r)_{0}$} &
\colhead{$r_{0}$} &
\colhead{$v_\odot$} &
\colhead{S/N} &
\colhead{Membership} &
\colhead{Photometry} \\
\colhead{} &
\colhead{(deg)} & 
\colhead{(deg)} &
\colhead{(mag)} &
\colhead{(mag)} &
\colhead{(km s$^{-1}$)} &
\colhead{(pixel$^{-1}$)} &
\colhead{} &
\colhead{}}
\startdata
1 & 336.17139 &  5.38661 &   0.47 &  22.28 & $-165.26\pm5.78$ & 1.89  &  N & IMACS \\
2 & 336.10198 &  5.38908 &   0.77 &  21.39 & $-220.57\pm4.71$ & 5.04 &  Y & IMACS\\
3 & 336.08657 &  5.39344 &  -0.08 &  22.05 & $-193.35\pm22.92$ & 1.43 &  ? & IMACS\\
4 & 336.09530 &  5.39583 &   0.37 &  21.27 & $-234.68\pm3.84$ & 4.32 &  Y & IMACS\\
5 & 336.11372 &  5.40772 &   0.49 &  20.88 & $-218.51\pm3.64$ & 7.00 &  Y & IMACS\\
6 & 336.09952 &  5.41176 &   0.65 &  21.07 & $-226.16\pm5.04$ &  5.73 &  Y & IMACS\\
7 & 336.11021 &  5.41590 &   0.53 &  20.94 & $-229.45\pm5.29$ &  6.56 &  Y & IMACS\\
8 & 336.07514 &  5.41965 &   0.63 &  21.65 & $-208.45\pm6.66$ &  4.10 &  Y & IMACS\\
9 & 336.09502 &  5.42197 &   0.58 &  21.94 &  $-50.69\pm6.94$ &  2.72 &   N & IMACS\\
10 & 336.10019 &  5.42418 &   0.68 &  21.07 & $-218.26\pm3.56$ & 6.37 & Y & IMACS\\
11 & 336.05841 &  5.43938 &   0.40 &  21.71 & $-260.11\pm9.97$ & 3.43 &  N & DECam\\
12 & 336.05740 &  5.45859 &   0.42 &  20.83 &  $-25.71\pm3.06$ &  6.13 &  N & IMACS
\enddata
\label{tab:TargetList}
\end{deluxetable*}

\section{Metallicity}

Given the low S/N of the spectra, we attempted to measure the spectra-averaged metallicity of the four brightest stars in our sample (\#5,6,7,10) using the Ca II triplet lines. The measured strength of the spectral lines for RGB stars can be calibrated to metallicity [Fe/H] based on the empirical relationship between the equivalent width and the luminosity offset from the HB of the system $V_0-V_{0,HB}$~\citep[e.g.,][]{Starkenburg2010, DaCosta2016}.  The $g_0$ and $r_0$ magnitudes of the stars were converted into $V_0$ magnitudes using the~\citep{Jordi2006} Pop II transformation equations. We calculated the $V_0-V_{0,HB}$ of the member stars based on the distance moduli for Peg\,III and M\,15 in Section 3, for which we assumed the $V_{HB}$ for M\,15 from Harris and E(B$-$V)=0.11 from~\cite{Kraft2003}. After smoothing the observed spectra with a 5-pixel boxcar using splot command in IRAF, we dealt with the spectra in two ways as follows. First, we add them all together in order to increase the S/N. The $\lambda8542\AA$ and $\lambda8662\AA$ Ca II line strengths were then measured using the procedure described in~\cite{DaCosta2016}. The summed equivalent width $\sum\,W\approx2.34\,\AA$  and the average $V_0-V_{0,HB}\approx-0.63$ mag give a reduced equivalent width $W'$ of $1.93\,\AA$. Applying the metallicity calibration with Equation (2) in~\cite{DaCosta2016} yields [Fe/H]=$-2.40$ dex with an uncertainty of order $0.15$ dex. We then added the spectra for the stars \#6 and \#10, and for \#5 and \#7 together separately. Repeating the above measurement process on these two spectra, we obtained $\sum\,W\approx2.72\,\AA$  with average $V_0-V_{0,HB}\approx-0.74$ mag for the star \#6 and \#10, and $\sum\,W\approx1.99\,\AA$ with average $V_0-V_{0,HB}\approx-0.51$ mag for the star \#5 and \#7. These values transform into [Fe/H]=$-2.24$ dex and [Fe/H]=$-2.55$ dex with uncertanties of order 0.15 dex, respectively. At face value, this is inconsistent with the CMD where the stars \#6 and \#10 appear $\sim0.2$ mag bluer than the stars \#5 and \#7 and so should be more metal-poor. The stars \#6 and \#10 are, however, possibly AGB stars to which the calibration process strictly may not apply. It must also be kept in mind that the S/N of the summed spectra, even after smoothing, remains low. Nevertheless, this result confirms that Peg\,III includes stars with metallicity as low as [Fe/H]=$-2.55\pm0.15$ dex.

\section{Kinematics}

\begin{figure}[t!]
\begin{centering}
\includegraphics[scale=0.49]{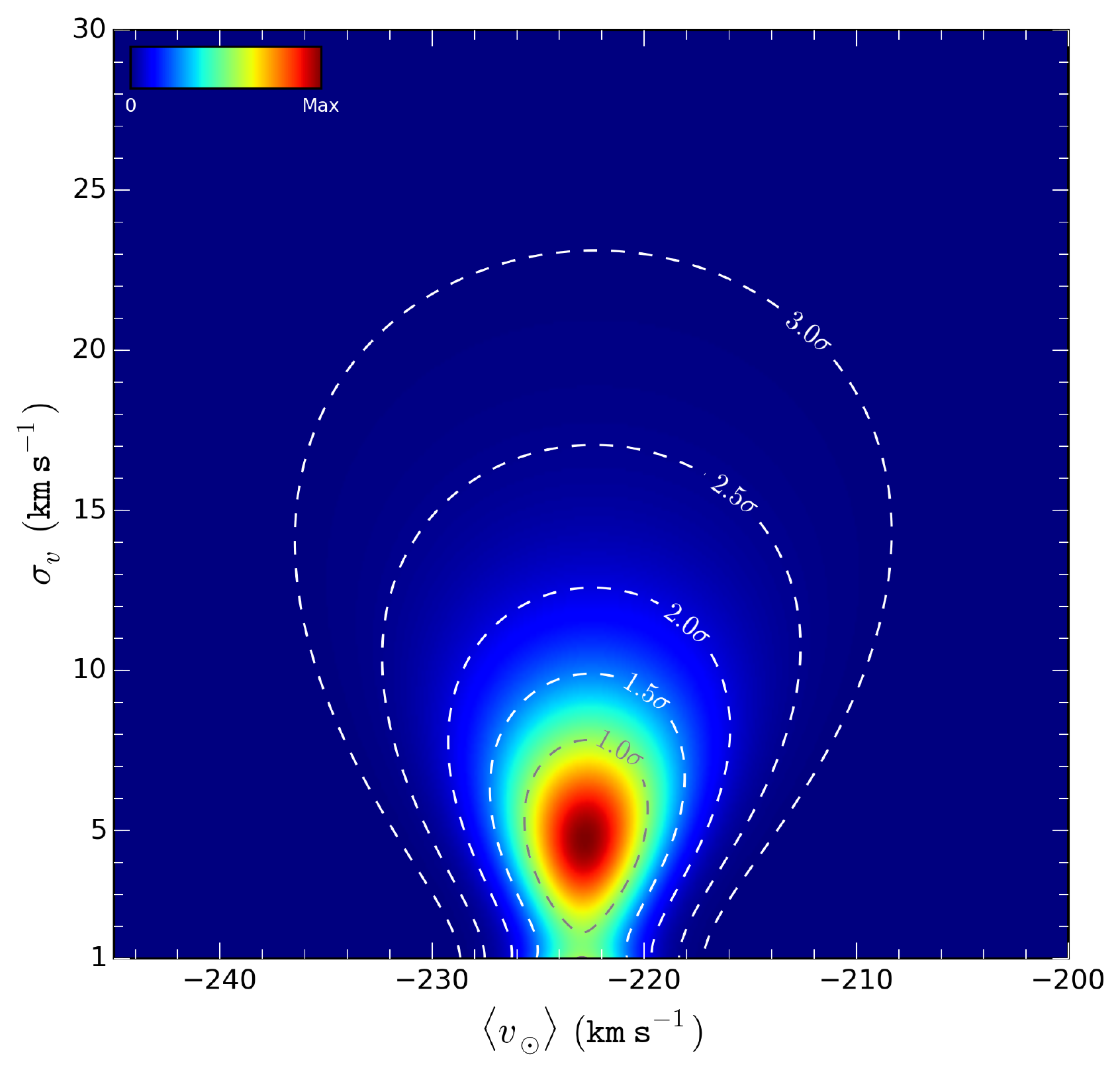}
\includegraphics[scale=0.49]{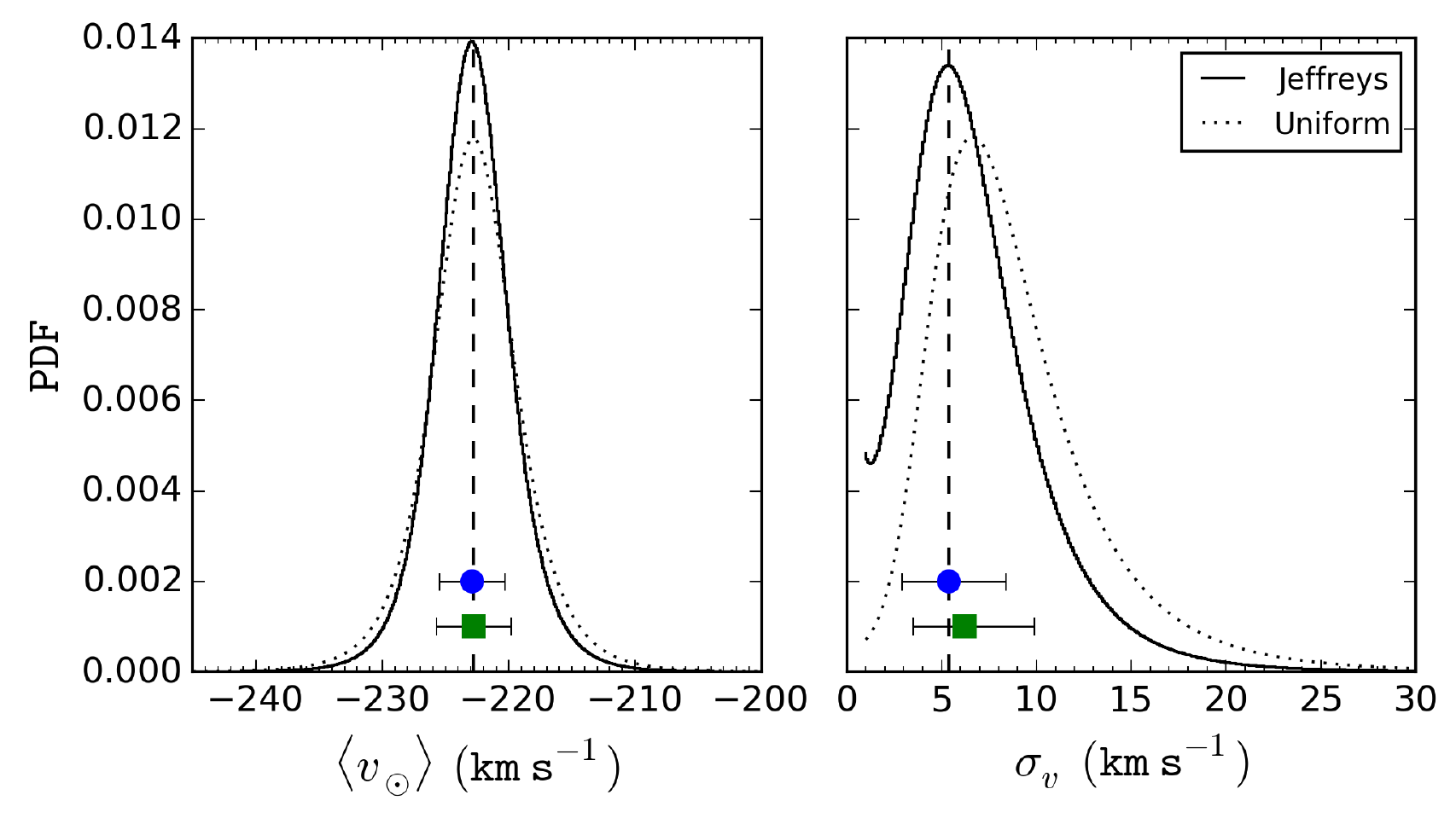}
\par\end{centering}
\protect\caption{Upper panel: two-dimensional posterior probability distribution for the mean velocity and velocity dispersion of Peg III. Contours outline the $1\sigma-3\sigma$ confidence levels.\footnote{In two dimensions, Gaussian densities within 1, 2, and 3$\sigma$ correspond to 39.3\%, 86.5\%, and 98.9\%, respectively.} Lower panels: corresponding marginalized PDFs (solid curves). The PDFs for a uniform prior on the velocity dispersion are overplotted for comparison (dotted curves). All the PDFs are normalized such that each PDF covers an equal probability density underneath. The dashed lines correspond to the modal values of the marginalized posterior PDFs. The circle and square with errorbars indicate the typical values and uncertainties determined by the method (a) and (b) in Section 5, respectively. \label{fig:Velocity}} 
\end{figure}

\begin{figure}[t!]
\begin{centering}
\includegraphics[scale=0.7]{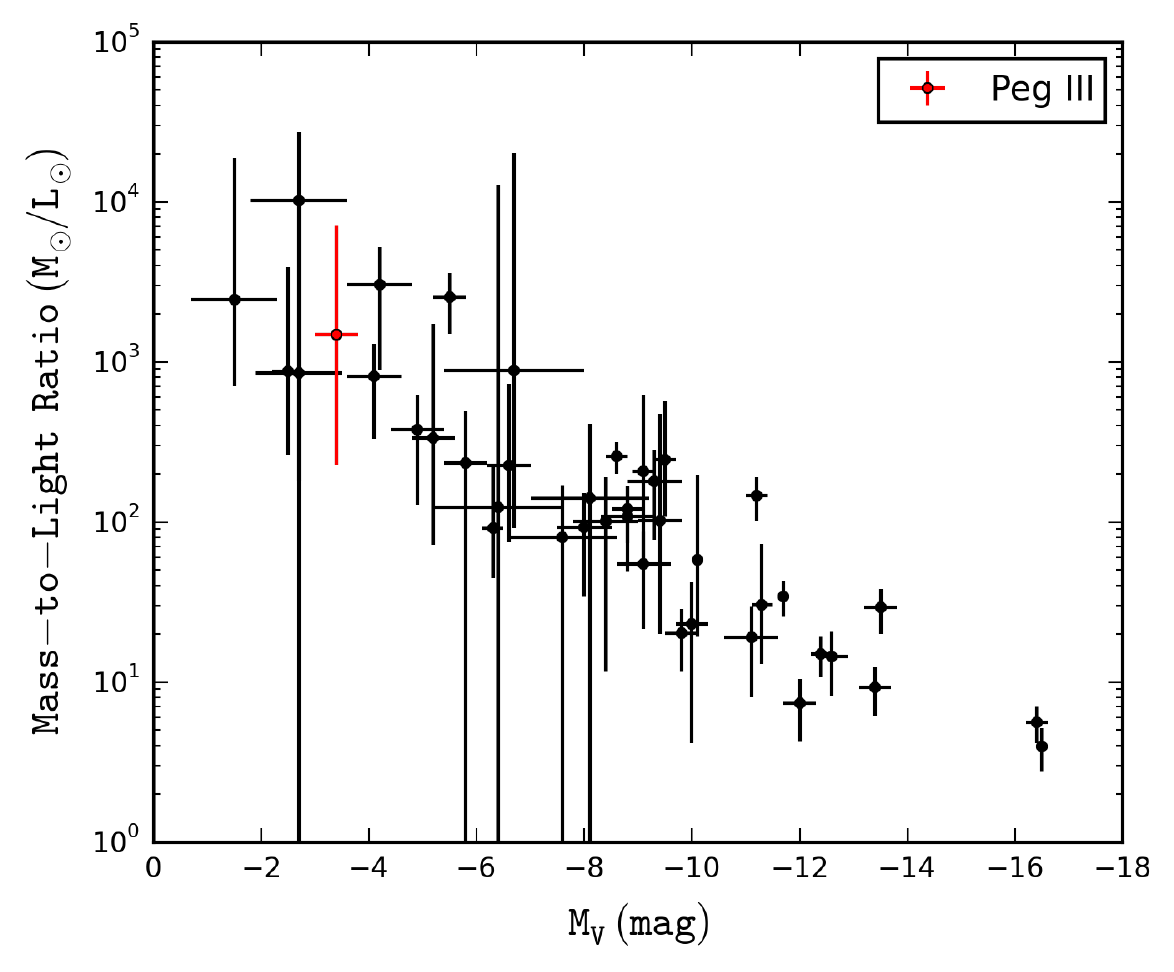}
\end{centering}
\protect\caption{ Mass-to-light ratio of Peg III (red) in comparison with other nearby galaxies within 1 Mpc. Mass-to-light ratios were calculated from the velocity dispersion, angular-sizes (half-light radii), heliocentric distances, and absolute magnitudes collected by \cite{McConnachie2012} for consistency with our estimate for Peg III. For the objects given ``symmetric'' uncertainties on the parameters, the errorbars were determined based on the regular error propagation, and for the rest based on the upper and lower limits of the parameters. \label{fig:MLratio}}
\end{figure}

In order to characterize the kinematics of Peg III, we employed a simple ``non-rotation'' model based on the method of~\cite{Drukier1998}. This method assumes that the measured radial velocities have a Gaussian distribution with mean velocity $\langle v_{\odot} \rangle$ and dispersion $\sigma_v$. The likelihood of the $i$th measurement $v_i\pm\delta_i$ is then given by

\begin{equation}
L_i(v_i\vert \langle v_{\odot} \rangle,\sigma_v)=G(v_{i}\vert \langle v_{\odot} \rangle, \sqrt{\sigma_v^2+\delta_i^2}),
\end{equation}

where $G(x\vert\mu,\sigma)$ is a Gaussian function of $x$ with the mean $\mu$ and the standard deviation $\sigma$. The likelihood for the available data set $D\equiv\lbrace v_i \rbrace_{i=1}^N$ is the product of the individual likelihoods:

\begin{equation}
L(D\vert \langle v_{\odot} \rangle,\sigma_v)=\prod\limits_{i}L_i(v_i\vert \langle v_{\odot} \rangle,\sigma_v).
\end{equation}

Applying Bayes' theorem leads to

\begin{equation}
P(\langle v_{\odot} \rangle,\sigma_v \vert D) \varpropto L(D\vert \langle v_{\odot} \rangle,\sigma_v)P(\langle v_{\odot} \rangle,\sigma_v),
\end{equation}

where $P(\langle v_{\odot} \rangle,\sigma_v \vert D)$ is the the posterior probability and $P(\langle v_{\odot} \rangle,\sigma_v)\equiv P(\langle v_{\odot} \rangle)P(\sigma_v)$ is the prior. For the mean velocity, the appropriate prior is a uniform prior $P(\langle v_{\odot} \rangle) = C$, for which we have set a finite range between $-200$ km s$^{-1}$ and $-245$ km s$^{-1}$ to make it normalizable. In the case of the velocity dispersion, the appropriate prior is the Jeffreys prior $P(\sigma_v)\varpropto\sigma_v^{-1}$~\citep[see, e.g.,][]{Drukier1998,Drukier2007,Koposov2015b,Torrealba2016b}, which is ``non-informative'' for a scale parameter such as the dispersion $\sigma_v$~\citep[see \textsection VII of][for justification]{Jaynes1968}. In fact, the choice of the prior has minimal impact on the posterior probability once the data are sufficiently constraining with a large sample size and small measurement errors. Otherwise, a uniform prior leads to a biased estimate for a scale parameter~\citep[see, e.g., \textsection 3.8 of][]{Gregory2005,Eriksen2008}. Since the Jeffreys prior $P(\sigma_v)\varpropto\sigma_v^{-1}$ is also an improper prior, it requires reasonable bounds to turn into a proper prior such that the likelihood distribution is not significantly truncated~\citep[see, e.g., \textsection 3.3 in][]{Drukier2007}. We have set a finite interval for the prior $\sigma_v\in(1,30)$ km s$^{-1}$, where the lower bound is $\sim1/5$ of the typical error on our measurements. We note that the likelihood for $\sigma_v\in(0,1)$ km s$^{-1}$ is equivalent to only 0.5\% of that for $\sigma_v\in(0,30)$ km s$^{-1}$.

 The upper panel of Figure~\ref{fig:Velocity} shows the resulting posterior probability distribution in two dimensional (2D) space, which appears asymmetric, spreading out toward larger velocity dispersions, most likely due to the small sample size~\citep[see, e.g., Figure 2 in][]{Walker2009}. The lower panels show the corresponding marginalized PDFs (solid curves) and also the PDFs constructed with a uniform prior on the velocity dispersion for comparison (dotted curves). Noticeably, the uniform prior favors larger velocity dispersions and displaces the modal value by $+1.2$ km s$^{-1}$. When it comes to determining the typical values and related uncertainties of the parameters, two different methods are commonly used in the literature; (a) find the modal values of the marginalised PDFs and the values that correspond to 61\% of the peak probability for the confidence interval~\citep[e.g.][]{Martin2014} or (b) read the 16, 50, and 84 percentiles of the marginalized PDFs~\citep[e.g.][]{Walker2015}. The results from each method are: (a) $\langle v_{\odot} \rangle=-222.9\pm 2.6$ km s$^{-1}$ and $\sigma_v=5.4^{+3.0}_{-2.5}$  km s$^{-1}$, and (b) $\langle v_{\odot} \rangle=-222.8^{+3.0}_{-2.9}$ km s$^{-1}$ and $\sigma_v=6.2^{+3.7}_{-2.7}$  km s$^{-1}$. We note that the inclusion of the ambiguous star\#3 with $v_\odot=-193.35\pm22.92$ km s$^{-1}$ in our sample does not make a significant difference in the results as follows: (a) $\langle v_{\odot} \rangle=-222.5\pm 2.6$ km s$^{-1}$ and $\sigma_v=5.4^{+3.1}_{-2.4}$  km s$^{-1}$, and (b) $\langle v_{\odot} \rangle=-222.3^{+3.1}_{-2.9}$ km s$^{-1}$ and $\sigma_v=6.3^{+3.8}_{-2.8}$  km s$^{-1}$. On the other hand, the exclusion of star \#4  with $v_\odot=-234.68\pm3.84$ km s$^{-1}$ from our sample leads to an unresolved solution for the velocity dispersion, no matter which one of the above two priors is used. We noticed the same phenomenon in a test with the member stars for Psc II reported by~\cite{Kirby2015a}; removing the star ID10694 with $v_\odot=-232.0\pm1.6$ km s$^{-1}$ causes an unresolved solution. In an experiment with the kinematic members of Segue 1 reported by~\cite{Simon2011}, we found that such an unresolved solution occurs in $\sim50$\% of the cases when 6 stars are randomly selected out of 32 stars having comparable measurement errors ($2<\delta_v<7$ km s$^{-1}$) and Bayesian membership probabilities larger than 90\%. This variation is even larger than the $1-\sigma$ uncertainty of the velocity dispersion and the influence of binary stars in the sample~\citep[$\sim0.5$ km s$^{-1}$, see Figure 6 in][]{Simon2011}. This result therefore suggests that the unresolved solutions are most likely a consequence of the small sample size. We will adopt the values and uncertainties obtained from method (a) as our final estimates in Table~\ref{tab:Parameters} and throughout the text. It is interesting to note that the measured systemic velocity for Peg\,III is very similar to that found for its neighbouring satellite Psc II ($\langle v_{\odot} \rangle=-226.5\pm 2.7$ km s$^{-1}$) independently measured by~\cite{Kirby2015a}.

Assuming dynamical equilibrium, the mass enclosed within the half-light radius of a stellar system can be accurately measured by the following equation as demonstrated by~\cite{Wolf2010}

\begin{equation}
\textup{M}_{1/2}\simeq\frac{4}{G}\,\sigma_v^{2}\,r_h\,  {\textup M}_{\odot},
\end{equation}

where $\sigma_v$ is the line-of-sight velocity dispersion and $r_h$ is the 2-dimensional projected half-light radius. According to this relation, the mass within the elliptical half light radius of Peg III is estimated to be $M_{1/2}=1.4^{+3.0}_{-1.1}\times10^6\, {\rm M}_{\odot}$. The absolute magnitude of Peg III we derived in Section 3, translates into a total luminosity of $1960\pm720\,{\rm L}_{\odot}$, which corresponds to a mass-to-light ratio of $ \textup{M/L}_V=1470^{+5660}_{-1240}\, {\rm M}_{\odot}/{\rm L}_{\odot}$. This value is consistent with the inverse correlation between luminosity and mass-to-light ratio for other nearby dwarf galaxies (see Figure~\ref{fig:MLratio}).

\section{Discussion and Summary}

We have obtained Magellan/IMACS photometry and Keck/DEIMOS spectroscopy for Peg III. The deep photometry confirms that Peg III is a faint ($M_{V}=-3.4\pm0.4$), elongated ($\epsilon=0.38^{+0.22}_{-0.38}$), irregular and distant ($d_{\odot}=215\pm12$ kpc) stellar system. We measured radial velocities for individual candidate member stars and identified seven, possibly eight member stars in the system based on their radial velocities, where the member stars could be either red giants or AGB stars (red large dots in Figure~\ref{fig:CMD}). The stellar population of Peg III contains stars with metallicity as low as [Fe/H]=$-2.55\pm0.15$ dex. The velocity dispersion of Peg III ($\sigma_v=5.4^{+3.0}_{-2.5}$ km s$^{-1}$) significantly exceeds the value expected from its observed stellar mass alone~\cite[$<0.3$ km s$^{-1}$; see Table 5 in][]{Pawlowski2015}, which supports the picture that Peg III is a satellite dwarf galaxy rather than a star cluster. 

Peg III and Psc II are approximately 43 kpc away from each other in three dimensions (3D) and their radial velocities in the Galactic standard-of-rest (GSR) frame differ only by $\sim10$ km s$^{-1}$ ($v_{GSR}=-67.6\pm2.6$ km s$^{-1}$ for Peg III and $v_{GSR}=-79.9\pm 2.7$ km s$^{-1}$ for Psc II). Given that only relatively few distant MW satellite galaxies are presently known, the close spatial proximity of Peg III and Psc II, and their very similar radial velocities, suggest a possible association between them. We note that another companionship of two distant MW satellites has been previously identified, namely the Leo IV - Leo V pair. Despite the difference in their radial velocities in the GSR frame~\citep[$\Delta v_{GSR}\sim50$ km s$^{-1}$, ][]{Simon2007,Belokurov2008}, their close spatial proximity in 3D ($\Delta d_{spatial}\sim22$\,kpc) has led to the hypothesis of a possible physical connection or common origin~\citep[e.g.][]{deJong2010}. Such a companionship of two satellites, as~\cite{Walker2009} suggested, may imply a rather circular orbit on which MW tides have a minimum effect. Assuming that the two satellites are currently a bound pair with an equal halo mass and follow a circular orbit on their average Galactocentric distance of $\sim198\pm10$\,kpc,  we have estimated their total halo mass following the method of~\cite{Evslin2014}. This method estimates the mass of the binary satellite system on the basis of the Virial theorem using the difference in their line-of-sight velocities ($12.3\pm3.7$\,km s$^{-1}$) and the separation between its constituents ($d_{spatial}=43\pm19$\,kpc) in 3D space. The derived mass of a satellite halo is $2.3\pm1.7\times10^{9}\/M_{\odot}$, which yields a tidal radius of $r_t=16\pm4$\,kpc. The ratio of the separation to the tidal radius is $d_{spatial}/r_t=2.7\pm1.0$. At face value, the tidal radius is smaller than the separation, in which case the binding energy of the two satellites is too low to remain undisrupted in the MW tidal field. This result, however, does not entirely rule out the possibility of physical pair as the tidal radius is comparable to the separation at the $1.7\sigma$ level. Information about their tangential velocities and dark-matter halo profiles, as well as more accurate measurements for other parameters, would provide more constraints on this result.

In both our DECam and IMACS photometry, Peg III appears irregular and elongated ($\epsilon$=$0.38^{+0.22}_{-0.38}$), at 1-$\sigma$ limit, compared to Psc II that only has an upper limit for its measured ellipticity~\citep[$\epsilon<0.28$;][]{Sand2012} with an unconstrained position angle. In fact, a similarity is found with the Leo IV - V pair~\citep[see Table 7 in][]{Sand2012}, where Leo V features larger ellipticity ($\epsilon$=$0.55\pm0.22$) with an unconstrained position angle and lower luminosity ($M_{V}=-4.4\pm0.4$) than Leo IV ($\epsilon<0.23$, $M_{V}=-5.5\pm0.3$). We consider three possible scenarios for the origin of the ellipticity of Peg III.  The first is that it is simply a result of its formation process.  The second is that it results from tidal interaction with the Milky Way.  Under the assumption that the stars of Peg III are in dynamic equilibrium this seems unlikely, given a) the large velocity dispersion implying a substantial mass, b) the compact physical size, c) the elongation misaligned with the direction toward the Galactic center, and d) the likelihood that Peg III and Psc II are moving on similar circular or near circular orbits at large Galactocentric distances.  On the other hand, if the large velocity dispersion of the Peg III stars reflects a non-equilibrium state or being inflated by unresolved binary stars rather than the presence of a large amount of dark matter, then the system might be a remnant of a dwarf galaxy tidally disrupted by the Milky Way.  This would require the orbit of Peg III to be significantly eccentric in order to reach the required smaller Galactocentric distances.  In turn, this would make the spatial and velocity agreement with Psc II coincidental, which seems highly unlikely. The third alternative is that the ellipticity of Peg III results from a tidal interaction with Psc II.  To test these different ideas will require N-body simulations or full 3D orbit information (i.e. the radial velocities and proper motions of the objects). Much deeper and wider imaging or higher resolution spectroscopy with new forthcoming telescopes such the Wide-Field Infrared Survey Telescope (WFIRST) and the Giant Magellan Telescope (GMT) may provide crucial keys to these questions.

\acknowledgements{
The authors thank the anonymous referee for helpful suggestions that improved the clarity and quality of the paper. We also thank Holger Baumgardt, Mila Chadayammuri, Joshua Simon, John Norris, and David Yong for interesting discussions and valuable comments. HJ and BC acknowledge the support of the Australian Research Council through Discovery projects DP120100475 and DP150100862. AC and AF acknowledge support from NSF-CAREER grant AST-1255160. APM acknowledges support by the Australian Research Council through Discovery Early Career Researcher Award DE150101816. GDC and ADM are grateful for support from the Australian Research Council through Discovery Projects DP120101237 and DP150103294. 

This research made use of Astropy, a community-developed core Python package for Astronomy~\citep{astropy}, and Matplotlib library~\citep{matplotlib}.

Some of the data presented herein were obtained at the W.M. Keck Observatory, which is operated as a scientific partnership among the California Institute of Technology, the University of California and the National Aeronautics and Space Administration. The Observatory was made possible by the generous financial support of the W.M. Keck Foundation. The authors wish to recognize and acknowledge the very significant cultural role and reverence that the summit of Mauna Kea has always had within the indigenous Hawaiian community.  We are most fortunate to have the opportunity to conduct observations from this mountain.}


\begin{thebibliography}{}

\bibitem[Ad{\'e}n et al.(2009)]{Aden2009} Ad{\'e}n, D., Feltzing, S., Koch, A., et al.\ 2009, \aap, 506, 1147 

\bibitem[Ahn et al.(2014)]{Ahn2014} Ahn, C.~P., Alexandroff, 
R., Allende Prieto, C., et al.\ 2014, \apjs, 211, 17 

\bibitem[Astropy Collaboration et 
al.(2013)]{astropy} Astropy Collaboration, Robitaille, T.~P., Tollerud, E.~J., et al.\ 2013, \aap, 558, A33 

\bibitem[Balbinot et al.(2013)]{Balbinot2013} Balbinot, E., 
Santiago, B.~X., da Costa, L., et al.\ 2013, \apj, 767, 101 

\bibitem[Bechtol et al.(2015)]{Bechtol2015} Bechtol, K., 
Drlica-Wagner, A., Balbinot, E., et al.\ 2015, \apj, 807, 50

\bibitem[Belokurov et al.(2006)]{Belokurov2006} Belokurov, V., 
Zucker, D.~B., Evans, N.~W., et al.\ 2006, \apjl, 647, L111

\bibitem[Belokurov et al.(2007)]{Belokurov2007} Belokurov, V., Zucker, D.~B., Evans, N.~W., et al.\ 2007, \apj, 654, 897 

\bibitem[Belokurov et al.(2008)]{Belokurov2008} Belokurov, V., Walker, M.~G., Evans, N.~W., et al.\ 2008, \apjl, 686, L83 

\bibitem[Belokurov et al.(2010)]{Belokurov2010} Belokurov, V., 
Walker, M.~G., Evans, N.~W., et al.\ 2010, \apjl, 712, L103

\bibitem[Belokurov et al.(2014)]{Belokurov2014} Belokurov, V., Irwin, 
M.~J., Koposov, S.~E., et al.\ 2014, \mnras, 441, 2124 

\bibitem[Bertin 
\& Arnouts(1996)]{SExtractor} Bertin, E., \& Arnouts, S.\ 1996, \aaps, 117, 393

\bibitem[Bertin et al.(2002)]{Swarp} Bertin, E., Mellier, Y., Radovich, M., et al.\ 2002, Astronomical Data Analysis Software and Systems XI, 281, 228 

\bibitem[Bertin(2006)]{Scamp} Bertin, E.\ 2006, Astronomical Data Analysis Software and Systems XV, 351, 112 

\bibitem[Bertin(2011)]{PSFEx} Bertin, E.\ 2011, Astronomical 
Data Analysis Software and Systems XX, 442, 435 

\bibitem[Bernard et al.(2014)]{Bernard2014} Bernard, E.~J., Ferguson, A.~M.~N., Schlafly, E.~F., et al.\ 2014, \mnras, 442, 2999  

\bibitem[Bond \& Neff(1969)]{Bond1969} Bond, H.~E., \& Neff, J.~S.\ 1969, \apj, 158, 1235 

\bibitem[Brown et al.(2014)]{Brown2014} Brown, T.~M., Tumlinson, J., Geha, M., et al.\ 2014, \apj, 796, 91 

\bibitem[Chabrier(2001)]{Chabrier2001} Chabrier, G.\ 2001, \apj, 
554, 1274 

\bibitem[Cooper et~al.(2012)]{Cooper2012} Cooper, M.~C., Newman, J.~A., Davis, M., Finkbeiner, D.~P., \& Gerke,
  B.~F. 2012, spec2d: DEEP2 DEIMOS Spectral Pipeline, Astrophysics Source Code Library

\bibitem[de Jong et al.(2010)]{deJong2010} de Jong, J.~T.~A., 
Martin, N.~F., Rix, H.-W., et al.\ 2010, \apj, 710, 1664

\bibitem[Da Costa(2016)]{DaCosta2016} Da Costa, G.~S.\ 2016, \mnras, 455, 199 

\bibitem[Desai et al.(2012)]{Desai2012} Desai, S., Armstrong, R., Mohr, J.~J., et al.\ 2012, \apj, 757, 83 

\bibitem[Dotter et al.(2008)]{Dartmouth} Dotter, A., Chaboyer, 
B., Jevremovi{\'c}, D., et al.\ 2008, \apjs, 178, 89 

\bibitem[Drlica-Wagner et al.(2015)]{DWagner2015} Drlica-Wagner, 
A., Bechtol, K., Rykoff, E.~S., et al.\ 2015, \apj, 813, 109 

\bibitem[Drukier et al.(1998)]{Drukier1998} Drukier, G.~A., Slavin, S.~D., Cohn, H.~N., et al.\ 1998, \aj, 115, 708 

\bibitem[Drukier et al.(2007)]{Drukier2007} Drukier, G.~A., Cohn, H.~N., Lugger, P.~M., et al.\ 2007, \aj, 133, 1041 

\bibitem[Eriksen et al.(2008)]{Eriksen2008} Eriksen, H.~K., Jewell, J.~B., Dickinson, C., et al.\ 2008, \apj, 676, 10-32 

\bibitem[Evslin(2014)]{Evslin2014} Evslin, J.\ 2014, \mnras, 440, 1225 

\bibitem[Faber et al.(2003)]{faber2003} Faber, S.~M., Phillips, A.~C., Kibrick, R.~I., et al.\ 2003, \procspie, 4841, 1657 

\bibitem[Fabrizio et al.(2014)]{Fabrizio2014} Fabrizio, M., Raimondo, G., Brocato, E., et al.\ 2014, \aap, 570, A61 

\bibitem[Gregory(2005)]{Gregory2005} Gregory, P.~C.\ 2005, Bayesian Logical Data Analysis for the Physical Sciences (Cambridge: Cambridge Univ. Press)

\bibitem[Gregory(2007)]{Gregory2007} Gregory, P.~C.\ 2007, \mnras, 374, 1321 

\bibitem[Hunter(2007)]{matplotlib} Hunter, J.~D.\ 2007, Computing 
in Science and Engineering, 9, 90 

\bibitem[Irwin et al.(2007)]{Irwin2007} Irwin, M.~J., Belokurov, 
V., Evans, N.~W., et al.\ 2007, \apjl, 656, L13 

\bibitem[Jaynes(1968)]{Jaynes1968} Jaynes, E. ~T. \ 1968, IEEE Trans. Syst. Sci. and Cybernetics, SSC-4(3), 227

\bibitem[Ji et al.(2016)]{Ji2016} Ji, A.~P., Frebel, A., Chiti, A., \& Simon, J.~D.\ 2016, \nat, 531, 610 

\bibitem[Jordi et al.(2006)]{Jordi2006} Jordi, K., Grebel, E.~K., \& Ammon, K.\ 2006, \aap, 460, 339 

\bibitem[Kraft \& Ivans(2003)]{Kraft2003} Kraft, R.~P., \& Ivans, I.~I.\ 2003, \pasp, 115, 143

\bibitem[Koposov et al.(2007)]{Koposov2007} Koposov, S., de Jong, 
J.~T.~A., Belokurov, V., et al.\ 2007, \apj, 669, 337

\bibitem[Koposov et al.(2015a)]{Koposov2015a} Koposov, S.~E., 
Belokurov, V., Torrealba, G., \& Evans, N.~W.\ 2015a, \apj, 805, 130 

\bibitem[Koposov et al.(2015b)]{Koposov2015b} Koposov, S.~E., Casey, 
A.~R., Belokurov, V., et al.\ 2015b, \apj, 811, 62 

\bibitem[Kim 
\& Jerjen(2015a)]{KimJerjen2015a} Kim, D., \& Jerjen, H.\ 2015a, \apj, 799, 73 

\bibitem[Kim et al.(2015a)]{Kim2015a} Kim, D., Jerjen, H., 
Milone, A.~P., Mackey, D., \& Da Costa, G.~S.\ 2015a, \apj, 803, 63

\bibitem[Kim et al.(2015b)]{Kim2015b} Kim, D., Jerjen, H., Mackey, D., Da Costa, G.~S., \& Milone, A.~P.\ 2015b, \apjl, 804, L44 

\bibitem[Kim et al.(2016)]{Kim2016} Kim, D., Jerjen, H., Mackey, D., Da Costa, G.~S., \& Milone, A.~P.\ 2016, \apj, 820, 119 

\bibitem[Kim 
\& Jerjen(2015b)]{KimJerjen2015b} Kim, D., \& Jerjen, H.\ 2015b, \apjl, 808, L39 

\bibitem[Kirby et al.(2013)]{Kirby2013} Kirby, E.~N., Cohen, J.~G., Guhathakurta, P., et al.\ 2013, \apj, 779, 102 

\bibitem[Kirby et al.(2015a)]{Kirby2015a} Kirby, E.~N., Simon, J.~D., \& Cohen, J.~G.\ 2015a, \apj, 810, 56

\bibitem[Kirby et al.(2015b)]{Kirby2015b} Kirby, E.~N., Cohen, 
J.~G., Simon, J.~D., \& Guhathakurta, P.\ 2015b, \apjl, 814, L7 

\bibitem[Kirby et al.(2015c)]{Kirby2015c} Kirby, E.~N., Guo, M., Zhang, A.~J., et al.\ 2015, \apj, 801, 125 

\bibitem[Laevens et al.(2014)]{Laevens2014} Laevens, B.~P.~M., 
Martin, N.~F., Sesar, B., et al.\ 2014, \apjl, 786, L3 

\bibitem[Laevens et al.(2015a)]{Laevens2015a} Laevens, B.~P.~M., Martin, N.~F., Ibata, R.~A., et al.\ 2015a, \apjl, 802, L18 

\bibitem[Laevens et al.(2015b)]{Laevens2015b} Laevens, B.~P.~M., 
Martin, N.~F., Bernard, E.~J., et al.\ 2015b, \apj, 813, 44 

\bibitem[Luque et al.(2016)]{Luque2016} Luque, E., Queiroz, A., Santiago, B., et al.\ 2016, \mnras, 458, 603 

\bibitem[Martin et al.(2008)]{Martin2008} Martin, N.~F., de Jong, 
J.~T.~A., \& Rix, H.-W.\ 2008, \apj, 684, 1075

\bibitem[Martin et al.(2014)]{Martin2014} Martin, N.~F., Chambers, K.~C., Collins, M.~L.~M., et al.\ 2014, \apjl, 793, L14 

\bibitem[Martin et al.(2015)]{Martin2015} Martin, N.~F., Nidever, 
D.~L., Besla, G., et al.\ 2015, \apjl, 804, L5

\bibitem[Martin et al.(2016a)]{martin2016a} Martin, N.~F., Geha, M., Ibata, R.~A., et al.\ 2016a, \mnras, 458, L59

\bibitem[Martin et al.(2016b)]{martin2016b} Martin, N.~F., Ibata, R.~A., Collins, M.~L.~M., et al.\ 2016b, \apj, 818, 40 

\bibitem[McConnachie(2012)]{McConnachie2012} McConnachie, A.~W.\ 2012, \aj, 144, 4

\bibitem[Mu{\~n}oz et al.(2012)]{Munoz2012} Mu{\~n}oz, R.~R., Padmanabhan, N., \& Geha, M.\ 2012, \apj, 745, 127

\bibitem[Newman et al.(2013)]{Newman2013} Newman, J.~A., Cooper, M.~C., Davis, M., et al.\ 2013, \apjs, 208, 5 

\bibitem[Robin et al.(2003)]{Robin2003} Robin, A.~C., Reyl{\'e}, C., Derri{\`e}re, S., \& Picaud, S.\ 2003, \aap, 409, 523 

\bibitem[Roederer et al.(2016)]{Raoederer2016} Roederer, I.~U., Mateo, M., Bailey, J.~I., III, et al.\ 2016, \aj, 151, 82 

\bibitem[Sand et al.(2009)]{Sand2009} Sand, D.~J., Olszewski, E.~W., Willman, B., et al.\ 2009, \apj, 704, 898 

\bibitem[Sand et al.(2010)]{Sand2010} Sand, D.~J., Seth, A., Olszewski, E.~W., et al.\ 2010, \apj, 718, 530 

\bibitem[Sand et al.(2012)]{Sand2012} Sand, D.~J., Strader, J., Willman, B., et al.\ 2012, \apj, 756, 79 

\bibitem[Schlafly \& Finkbeiner(2011)]{Schlafly2011} Schlafly, E.~F., \& Finkbeiner, D.~P.\ 2011, \apj, 737, 103 

\bibitem[Schlegel et al.(1998)]{Schlegel1998} Schlegel, D.~J., Finkbeiner, D.~P., \& Davis, M.\ 1998, \apj, 500, 525

\bibitem[Simon \& Geha(2007)]{Simon2007} Simon, J.~D., \& Geha, M.\ 2007, \apj, 670, 313 

\bibitem[Simon et al.(2011)]{Simon2011} Simon, J.~D., Geha, M., Minor, Q.~E., et al.\ 2011, \apj, 733, 46 

\bibitem[Simon et al.(2015)]{Simon2015} Simon, J.~D., 
Drlica-Wagner, A., Li, T.~S., et al.\ 2015, \apj, 808, 95 

\bibitem[Starkenburg et al.(2010)]{Starkenburg2010} Starkenburg, E., Hill, V., Tolstoy, E., et al.\ 2010, \aap, 513, A34 

\bibitem[Taylor(2005)]{STILTS} Taylor, M.~B.\ 2005, Astronomical Data Analysis Software and Systems XIV, 347, 29 

\bibitem[Tonry et al.(2012)]{Tonry2012} Tonry, J.~L., Stubbs, C.~W., Lykke, K.~R., et al.\ 2012, \apj, 750, 99 

\bibitem[Torrealba et al.(2016a)]{Torrealba2016a} Torrealba, G., Koposov, S.~E., Belokurov, V., \& Irwin, M.\ 2016a, \mnras, 459, 2370

\bibitem[Torrealba et al.(2016b)]{Torrealba2016b} Torrealba, G., Koposov, S.~E., Belokurov, V., et al.\ 2016b, \mnras, 463, 712 

\bibitem[Pawlowski et al.(2015)]{Pawlowski2015} Pawlowski, M.~S., McGaugh, S.~S., \& Jerjen, H.\ 2015, \mnras, 453, 1047

\bibitem[Vargas et al.(2013)]{Vargas2013} Vargas, L.~C., Geha, M., Kirby, E.~N., \& Simon, J.~D.\ 2013, \apj, 767, 134 

\bibitem[Voggel et al.(2016)]{Voggel2016} Voggel, K., Hilker, M., Baumgardt, H., et al.\ 2016, \mnras, 460, 3384 

\bibitem[Walker et al.(2009)]{Walker2009} Walker, M.~G., Belokurov, V., Evans, N.~W., et al.\ 2009, \apjl, 694, L144

\bibitem[Walker et al.(2015)]{Walker2015} Walker, M.~G., Mateo, 
M., Olszewski, E.~W., et al.\ 2015, \apj, 808, 108 

\bibitem[Walker et al.(2016)]{Walker2016} Walker, M.~G., Mateo, M., Olszewski, E.~W., et al.\ 2016, \apj, 819, 53 

\bibitem[Walsh et al.(2007)]{Walsh2007} Walsh, S.~M., Jerjen, H., 
\& Willman, B.\ 2007, \apjl, 662, L83 

\bibitem[Walsh et al.(2008)]{Walsh2008} Walsh, S.~M., Willman, B., Sand, D., et al.\ 2008, \apj, 688, 245-253 

\bibitem[Weisz et al.(2016)]{Weisz2016} Weisz, D.~R., Koposov, S.~E., Dolphin, A.~E., et al.\ 2016, \apj, 822, 32 

\bibitem[Willman et al.(2005)]{Willman2005} Willman, B., Blanton, 
M.~R., West, A.~A., et al.\ 2005, \aj, 129, 2692 

\bibitem[Willman \& Strader(2012)]{Willman2012} Willman, B., \& Strader, J.\ 2012, \aj, 144, 76 

\bibitem[Wolf et al.(2010)]{Wolf2010} Wolf, J., Martinez, G.~D., Bullock, J.~S., et al.\ 2010, \mnras, 406, 1220 

\bibitem[York et al.(2000)]{York2000} York, D.~G., Adelman, J., 
Anderson, J.~E., Jr., et al.\ 2000, \aj, 120, 1579 

\bibitem[Zucker et al.(2006)]{Zucker2006} Zucker, D.~B., 
Belokurov, V., Evans, N.~W., et al.\ 2006, \apjl, 643, L103 

\end{thebibliography}

\end{document}